\documentclass[twocolumn,showpacs,preprintnumbers,amsmath,amssymb]{revtex4}

\usepackage{graphicx}
\usepackage{dcolumn}
\usepackage{bm}

\begin{document}
\title{Sagnac Interferometry in a Slow-Light Medium}

\author{Graham T. Purves}
\email{g.t.purves@dur.ac.uk}
\homepage{http://massey.dur.ac.uk/gtp}
\author{Charles S. Adams}
\author{Ifan G. Hughes}
\affiliation{Department of Physics, Durham University, South Road, Durham, DH1 3LE, UK.}

\date{\today}

\begin{abstract}
We use a Sagnac interferometer to measure the dispersive and absorptive properties of room temperature Rubidium vapor on the D$_{2}$ line at $780.2$ nm.  We apply a pump beam such that the resulting $\Lambda$ system exhibits Electromagnetically Induced Transparency.  Using a ``biased alignment" technique we demonstrate a direct and robust method of measuring the rapid variation in the refractive index.  Such a ``slow-light" Sagnac interferometer is ideally suited to precision measurement applications such as magnetometry and inertial sensing.\end{abstract}

\maketitle
Slow light is the propagation of light with a group velocity significantly below the speed of light in a vacuum.  Slow light propagation is observed in media whose refractive index varies rapidly with respect to frequency.  The rapid variation in the refractive index has a concomitant narrow transparent window in the spectrum of the slow-light medium.  Such narrow transmission windows can be produced using Electromagnetically Induced Transparency (EIT) in an atomic vapor \cite{Hau:nature1999,Kash:prl1999,Budker:prl1999}.  Slow light propagation has also been observed in solid-state media such as ruby \cite{Bigelow:prl2003}
and optical fibers \cite{Okawachi:prl2005}.
In general EIT is an effect seen in three-level systems interacting with two light fields \cite{Harris:prl1990,Boller:prl1991}.  For slow light in atomic systems, the three levels form a $\Lambda$ system with the two lower states being coupled to the common upper state by optical transitions.  In the $\Lambda$ system the width of the EIT resonance is ultimately limited by the coherence time between the two lower energy states, which can be very long, consequently resonances as narrow as tens of Hz have been observed \cite{Brandt:pra1997,Erhard:pra2001}.
Narrow resonances lend themselves to being used as the basis for sensitive measurements.  For example, utilising the narrow linewidth of EIT resonances, it is possible to construct extremely compact atomic clocks \cite{Knappe:apl2004}.  Magnetically sensitive $\Lambda$ systems enable the construction of compact and sensitive magnetometers \cite{Schwindt:apl2004,Scully:prl1992,Nagel:epl1998,Stahler:epl2001,Affolderbach:apb2002}.
It has also been proposed that slow light could enhance the rotational sensitivity of a Sagnac interferometer \cite{Sagnac:cras1913}, to produce a hybrid optical matter-wave Sagnac interferometer \cite{Zimmer:prl2004} --- this scheme would require laser cooled atoms to achieve the proposed matter-wave sensitivity gain.  The sensitivity of atomic systems to inertial sensing is also highlighted in recent work where an atomic magnetometer has been used to provide precision measurements of rotations \cite{Kornack:prl2005}.

\begin{figure}[ht]\centering
\includegraphics[height=7.0cm,angle=0]{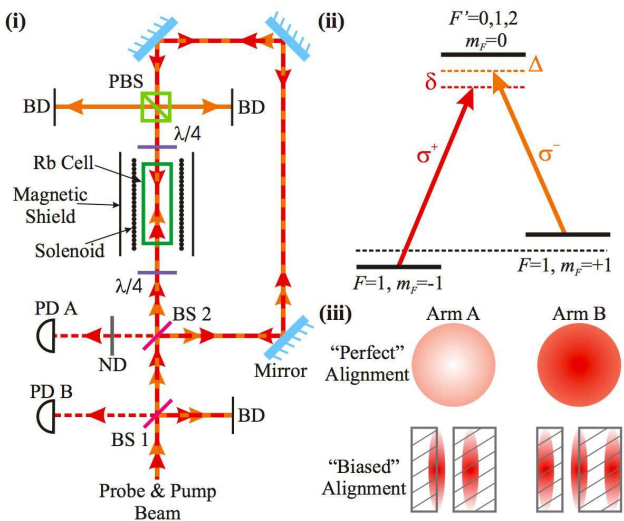}
\caption{\label{fig:sagnac} (Color online) (i) Sagnac interferometer experimental set-up.  The probe beam (red/dark-grey) passes around the Sagnac interferometer in both clockwise and anticlockwise directions, however the orthogonally polarized pump beam (orange/light-grey) only passes through the Rubidium vapor cell in the clockwise direction.  BS --- 50:50 beam splitter; ND --- neutral-density filter; PD --- photodiode; PBS --- polarization beam spitting cube; BD --- beam dump; and $\lambda$/4 --- quarter-wave plate.  (ii) The $\Lambda$ system investigated in this paper: $\Delta$ is the single-photon detuning of the pump field from the atomic resonance; and $\delta$ is the detuning of the probe from the two-photon Raman resonance.  (iii)  Output beam profiles for ``perfect" alignment (upper) and ``biased" alignment (lower), see text.  The shaded rectangles show the positions of two mechanical slits. }
\end{figure}

In this paper, we present measurements on a slow-light medium in a Sagnac interferometer. We show that an appropriately configured slow-light Sagnac interferometer provides a direct and robust readout of the dispersion. The experimental set-up is shown in Fig. \ref{fig:sagnac}(i).
We study the $^{87}$Rb D$_{2}$ line ($5s\  ^{2}S_{\frac{1}{2}}\ F=1 \rightarrow 5p\ ^{2}P_{\frac{3}{2}}\ F'$) using orthogonal circular polarized pump and probe beams. For a room temperature vapor, the pump and probe couple to three excited states, as depicted in Fig. \ref{fig:sagnac}(ii).

The degeneracy of the two Zeeman sub-levels within the ground hyperfine state is lifted by applying a magnetic field of $\simeq 1$ G co-axial to the pump and probe beams.  This ensures that the two-photon resonance ($\delta = 0$) does not coincide with the beat signal arising when the pump and probe have the same frequency.

Both pump and probe beams are derived from the same Extended Cavity Diode Laser (ECDL) and double-pass through separate Acousto-Optic Modulators (AOMs). The ECDL is detuned by $\Delta$ from the $F=1 \rightarrow F'=2$ resonance.  The pump beam is shifted by a fixed frequency offset, whereas the probe has a variable offset which is scanned about $\delta = 0$, see Fig. \ref{fig:sagnac}(ii).  The frequency-shifted beams are recombined on a polarizing beam splitter (PBS) and coupled through a single mode, polarization-preserving optical fiber ensuring that the pump and probe beams are perfectly co-propagating. The output from the fiber is
collimated and passes through two 50:50 beam splitters. The first 
beam splitter is used to pick off one of the Sagnac outputs.
The loss associated with the first 50:50 beam splitter could be avoided
by using polarization to pick off the output beam \cite{Jundt:epjd2003}, however this would prevent the use of orthogonally polarized pump and probe beams.
The second beam splitter divides both pump and probe into the two counter-propagating arms of the Sagnac interferometer, see Fig. \ref{fig:sagnac}(i).
The beams within the Sagnac interferometer all have circular Gaussian profiles, with a full-width at half-maximum (FWHM) in the center of the Rb vapor cell of around $2$ mm.  Either side of the Rb cell, the orthogonal linear polarized beams are converted into circular polarizations by quarter-wave plates.  The beams pass through an 8 cm long Rb vapor cell, which is positioned centrally within a solenoid, inside a cylindrical $\mu$-metal shield.
In the anticlockwise direction, the pump beam is coupled out of the interferometer by a PBS, before reaching the Rb cell.
The two counter-propagating probe beams interfere on the second beam splitter.  Output arm A includes a neutral-density filter which transmits $50\%$ of the light onto photodiode A.  Arm B propagates back towards the first beam splitter, where $50\%$ is reflected onto photodiode B.  

In the case that the probe beams perfectly counter-propagate, (``Perfect" alignment, Fig. \ref{fig:sagnac}(iii)(upper)); arm A is dark (light grey/light red), whilst arm B is bright (dark grey/red). A change in the refractive index of the medium for one direction of propagation shifts the fringe pattern.  However the sensitivity is minimal, as the shift is about a maxima or minima of the interference pattern, where the rate of change of intensity with displacement is lowest.  To enhance the sensitivity we ``bias" the interferometer by introducing a small angle between the counter-propagating beams \cite{Robins:ol2002,Jundt:epjd2003}, such that both light and dark fringes appear in the interference pattern at both outputs, Fig. \ref{fig:sagnac}(iii)(lower).  Two mechanical slits aperture the fringe pattern, such that only the region between the light and dark fringe is focussed onto the photodiode.  This biasing technique enables one to obtain maximal sensitivity to changes in the refractive index and a signal that is directly proportional to the refractive index difference between the two counter-propagating probes.  The sum and difference signal between the two outputs are given by
\begin{equation}
 \frac{P_{\textrm{A}}+P_{\textrm{B}}}{P_{0}} \propto \left( \textrm{e}^{-\alpha_{\textrm{c}}L}+\textrm{e}^{-\alpha_{\textrm{a}}L} \right)~,
\label{eqn:1}
\end{equation}
\begin{equation}
\frac{P_{\textrm{A}}-P_{\textrm{B}}}{P_{0}} \propto \textrm{e}^{-\alpha L}\sin \left( kL\Delta n \right)~,
\label{eqn:2}
\end{equation}
where $P_{\textrm{A}}$, $P_{\textrm{B}}$ and $P_{0}$ quantify the power in arms A, B and the total power into the interferometer, respectively;  $\alpha_{\textrm{c}}$ and $\alpha_{\textrm{a}}$ are the absorption coefficients for the clockwise and anticlockwise beams, respectively; $\alpha = \left( \alpha_{\textrm{c}} + \alpha_{\textrm{a}}\right) / 2$;  $L$ is the length of  the Rb vapor cell; and $\Delta n$ is the difference in refractive index between the clockwise and anticlockwise beams.  It should be noted that $\vert kL \Delta n \vert \ll 1$, hence $\sin (kL \Delta n) \simeq kL \Delta n$, i.e. the difference signal can be taken to be directly proportional to $\Delta n$.
Finally, as only the clockwise probe beam co-propagates with the pump beam within the vapor cell, only the clockwise probe beam experiences a narrow EIT feature, hence $\Delta n$ is equal to the refractive index of the EIT feature alone.  Thus the difference between the two output arms of the Sagnac interferometer is directly proportional to the dispersion associated with the EIT feature. 

Compared to a Mach-Zehnder interferometer \cite{Xiao:prl1995,Zibrov:prl1996},  the Sagnac interferometer provides a more robust tool for measuring EIT dispersion features.  In both interferometers it is necessary that the beam paths of the two arms should be close to equal.  This is trivial in the Sagnac, as both arms of  the interferometer follow the same path but in opposite directions, whereas in the Mach-Zehnder the path length of the two independent arms must be matched (requiring a piezoelectric transducer to control of the length of one of the two arms \cite{Zibrov:prl1996}).  In addition the Sagnac interferometer provides a degree of common-mode rejection against translational mirror vibrations.

\begin{figure}[ht]\centering
\includegraphics[height=10cm,angle=0]{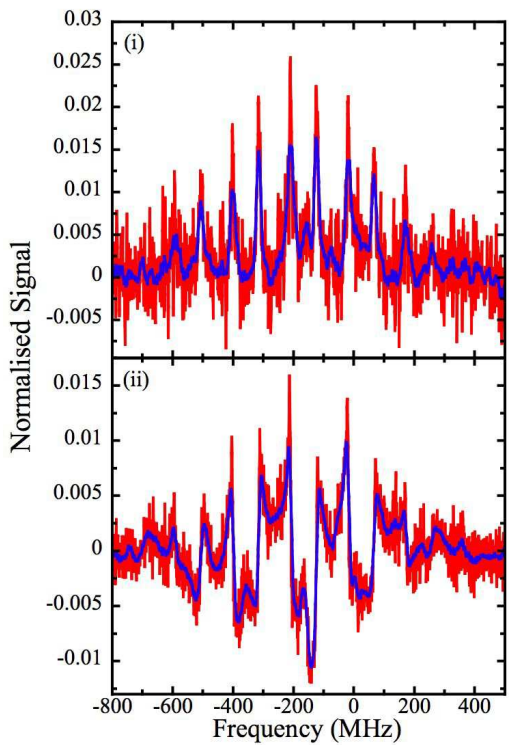}
\caption{\label{fig:double_scan} (Color online) Sum (i) and difference (ii) of the output arms of the Sagnac interferometer.   The red (light-grey) trace shows unsmoothed data, the blue (dark-grey) trace shows the data with a nineteen point running average.   The probe beam power is $4\ \mu$W, the pump beam power is $26\ \mu$W.  The piezo  is used to scan both pump and probe fields over a range of 1.3 GHz whilst the AOM scans the probe field twelve times, over a range of 1.8 MHz, about the two-photon Raman resonance.  The frequency scale applies only to the piezo scan.  When the Raman resonance condition is met, there is a sharp increase in transmission and a concomitant rapid change in the refractive index.  The signals in both plots are normalised such that the sum of the signals in the two arms, away from the Doppler resonance, is equal to one.  The Doppler broadened background has been subtracted from both signals. }
\end{figure}

To ascertain at which value of laser detuning, $\Delta$, the features are strongest, a double-scanning technique was used \cite{Akulshin:job2003,Akulshin:pra2003}.  $\Delta$ is scanned over the full Doppler broadened transition once, whilst at the same time, the two-photon detuning $\delta / 2 \pi$, is scanned $m$ times over a range of several hundreds of kHz with the AOM \cite{Akulshin:job2003,Akulshin:pra2003}.  This leads to $m$ EIT features occurring within the range of the Doppler broadened transition Fig. \ref{fig:double_scan}.  The frequency scale of such a plot is not straight forward, the centers of the different EIT features are separated by a frequency given by the ECDL scan $\Delta / 2\pi$ whilst the width of the features is determined by the AOM scan $\delta / 2 \pi$.
The amplitude of the EIT signals are determined by a Gaussian envelope.  This is expected, as the number of atoms that contribute to the EIT signal across the Doppler broadened resonance can be closely approximated by a Gaussian function.  Scanning two counter-propagating beams at the same frequency across the Doppler broadened resonance leads to the occurrence of saturation spectroscopy resonances \cite{MacAdam:ajp1992,Smith:ajp2004}.  The features most prominent in Fig. \ref{fig:double_scan} occur at frequencies of approximately $0$~MHz, $-80$~MHz and $-160$~MHz, consistent with the known hyperfine structure.

\begin{figure}[ht]\centering
\includegraphics[height=12cm,angle=0]{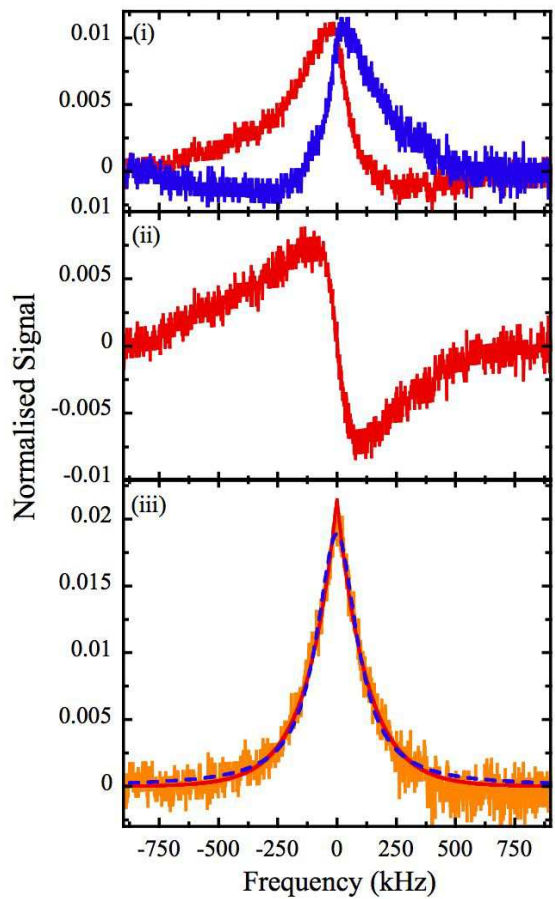}
\caption{\label{fig:single_scan} (Color online) The EIT feature as measured in the Sagnac interferometer. The probe power is $3\ \mu$W and the pump power is $28\ \mu$W.  The signal size is normalised to the sum of the off resonance signal in both arms.  The background signal has been subtracted from each arm after normalisation.  (i) shows the individual signals from the two output arms of the Sagnac interferometer: A --- red (light-grey) and B --- blue (dark-grey).  (ii) shows the difference signal between the two arms A $-$ B.  (iii) shows the sum signal of the two arms (orange/light-grey line), with least-square fits of a Lorentzian function (blue/dark-grey dashed line) and a cusp function (red/mid-grey solid line).}
\end{figure}

\begin{figure}\centering
\includegraphics[height=6cm,angle=0]{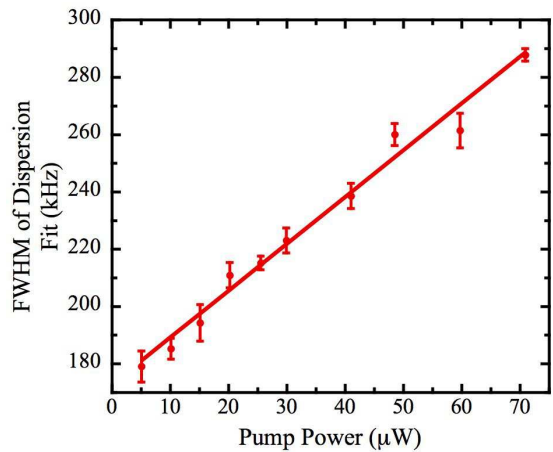}
\caption{\label{fig:widths} Plot of the FWHM of a Lorentzian dispersion fit to the difference signal for a range of pump powers. Experimental data (circles) with a linear fit (solid line).}
\end{figure}

In order to characterize both the transmission and dispersion of the EIT feature, the ECDL was tuned to the center of the Doppler broadened transition, between $-100$ and $-200$~MHz in Fig. \ref{fig:double_scan}.  The probe alone was then scanned across the two-photon Raman resonance.  The signals from the two output arms, A and B, were recorded, both with and without the pump field.  The probe-only signal was subtracted from the probe-and-pump signal, this removed varying background due to the scanning of the AOM.   Plots of the two individual arms of a typical signal can be seen in Fig. \ref{fig:single_scan}(i).  The signal is normalised to the off-resonance sum of the two output arms, and then the background is subtracted.
The difference signal, $P_{\textrm{A}} - P_{\textrm{B}}$ (Eq.~(\ref{eqn:2})), giving the refractive index of the EIT feature, is shown in Fig.~\ref{fig:single_scan}(ii).  The sum Eq.~(\ref{eqn:1}), which is proportional to the transmission of the EIT resonance, is shown in Fig.~\ref{fig:single_scan}(iii).  Along with the sum signal data are two least-square fits, a Lorentzian and a cusp function (two back-to-back exponentials).  Both functions are fitted to the data across the frequency range shown in the figure.  If the signals are power broadened, the beam profile affects the line shape and width of the resonance \cite{Taichenachev:pra2004}: a step-like beam profile leads to a Lorentzian line shape and a Gaussian beam profile to an arctan line shape.  In this case the FWHM varies linearly with the intensity of the beam, Fig.~\ref{fig:widths}.  The cusp function gives the expected line shape for transit-time dominated broadening \cite{Thomas:pra1980}, and is virtually indistinguishable from the arctan fit \cite{Taichenachev:pra2004}.

For the parameters used in this work, the EIT resonance is power broadened.  The amplitude of the features increases with pump power until it starts to saturate at around $50\ \mu$W.  Reducing the intensity of the pump reduces the width of the resonances, as can be seen in Fig.~\ref{fig:widths}.
Extrapolating the linear fit of Fig.~\ref{fig:widths}, shows that reducing the pump power to zero will lead to a FWHM of 170 kHz, where the probe power is $4\ \mu$W.  Whilst this is more than forty times narrower than the excited state linewidth, even narrower resonances could be obtained using a buffer gas to extend the interaction time \cite{Brandt:pra1997,Erhard:pra2001}.
For example by increasing the beam size by a factor of two we obtain resonance widths of 70 kHz. A larger beam size has a two-fold benefit: the intensity is reduced and the transit time of atoms through the pump beam is increased --- leading to reduced transit time broadening.

In summary, a Sagnac interferometer has been realised incorporating a slow-light medium. We show that one can ``bias" the interferometer to provide a direct and robust readout of the dispersion associated with narrow EIT resonances. Using this technique Sagnac interferometry provides a modulation-free method of producing a signal with potential applications in magnetometry and inertial sensing. 
\begin{acknowledgments}
Graham Purves's PhD studentship was provided by EPSRC. the authors would like to thank M. Pritchard, Dr. S. Cornish and Dr. S. Gardiner for useful comments on the text.
\end{acknowledgments}
\appendix

\end{document}